\documentclass[5p, authoryear]{elsarticle}
\usepackage{epsfig,subfigure,amsmath}
\usepackage{color}
\usepackage{hyperref}

\definecolor{purple}{rgb}{1,0,1}

\newcommand{\lcdm}{$\Lambda$CDM}
\newcommand{\hmpc}{$h^{-1}$Mpc}

\newcommand{\beq}{\begin{equation}}
\newcommand{\eeq}{\end{equation}}
\def\vide{\textsc{vide}}
\def\healpix{{\tt healpix}}

\DeclareMathOperator\erf{erf}
\def\ap{Alcock-Paczy\'{n}ski}

\bibpunct[; ]{(}{)}{;}{a}{}{,}

\def\apjl{ApJ}

\def\apjs{ Astrophys. J. Supp.}

\def\aap{Astron. \& Astrophys.}

\begin{document}

\begin{frontmatter}

\title{VIDE: The Void IDentification and Examination Toolkit}

\author[add1,add2,add3]{P.~M.~Sutter}
\ead{sutter@iap.fr}
\author[add1,add2]{Guilhem Lavaux}
\author[add1,add2]{Nico Hamaus}
\author[add1,add2]{Alice Pisani}
\author[add1,add2,add4,add5]{Benjamin D. Wandelt}
\author[add6]{Mike Warren}
\author[add7,add8]{Francisco Villaescusa-Navarro}
\author[add9]{Paul Zivick}
\author[add10]{Qingqing Mao}
\author[add11]{Benjamin B. Thompson}

\address[add1]{Sorbonne Universit\'{e}s, UPMC Univ Paris 06, UMR7095, Institut d'Astrophysique de Paris, F-75014, Paris, France}
 \address[add2]{CNRS, UMR7095, Institut d'Astrophysique de Paris, F-75014, Paris, France}
 \address[add3]{Center for Cosmology and AstroParticle Physics, Ohio State University, Columbus, OH 43210, USA}
 \address[add4]{Department of Physics, University of Illinois at Urbana-Champaign, Urbana, IL 61801, USA}
 \address[add5]{Department of Astronomy, University of Illinois at Urbana-Champaign, Urbana, IL 61801, USA}
 \address[add6]{Theoretical Division, Los Alamos National Laboratory, Los Alamos, NM 87545, USA}
 \address[add7]{INAF - Osservatorio Astronomico di Trieste, Via Tiepolo 11, 1-34143, Trieste, Italy}
 \address[add8]{INFN sez. Trieste, Via Valerio 2, 1-34127, Trieste, Italy}
 \address[add9]{Department of Astronomy, Ohio State University, Columbus, OH 43210, USA}
 \address[add10]{Department of Physics and Astronomy, Vanderbilt University, Nashville, TN 37235, USA}
 \address[add11]{Jeremiah Horrocks Institute, University of Central Lancashire, Preston, PR1 2HE, United Kingdom}

\begin{abstract}
We present \vide, the Void IDentification and Examination toolkit, an
open-source Python/C++ code for finding cosmic voids in galaxy redshift surveys 
and $N$-body simulations, characterizing their properties, and providing
a platform for more detailed analysis.
At its core, \vide~uses a substantially enhanced version of 
\textsc{zobov} (Neyinck 2008) to calculate a 
Voronoi tessellation 
for estimating the density field and a performing a 
watershed transform to construct 
voids. 
Additionally, \vide~provides significant functionality for both pre- and 
post-processing: for example, \vide~can work with volume- or 
magnitude-limited galaxy samples with arbitrary survey geometries, 
or dark 
matter particles or halo catalogs in a variety of common formats.
It can also randomly subsample inputs
and includes a Halo Occupation Distribution model for 
constructing mock galaxy populations.
\vide~uses the watershed levels to place voids in a hierarchical tree,
outputs a summary of void properties in plain ASCII, and provides 
a Python API to perform many analysis tasks, such as loading and 
manipulating void catalogs and particle members, filtering, plotting, 
computing clustering statistics, stacking, comparing catalogs, and 
fitting density profiles.
While centered around \textsc{ZOBOV}, the toolkit is designed to be 
as modular as possible and accommodate other void finders.
\vide~has been in development for several years and has already been used 
to produce a wealth of results, which we summarize in this work to 
highlight the capabilities of the toolkit.
\vide~is publicly available at  
{http://bitbucket.org/cosmicvoids/vide\_public}
and
\mbox{http://www.cosmicvoids.net}.
\end{abstract}


\end{frontmatter}


\section{Introduction}

Cosmic voids are emerging as a novel probe of both 
cosmology and astrophysics, as well as fascinating objects of study 
themselves. These large empty regions in the cosmic 
web, first discovered over thirty years ago~\citep{Gregory1978,Joeveer1978,Kirshner1981}, 
are now known to fill up nearly the entire volume of the 
Universe~\citep{Hoyle2004,Pan2011,Sutter2012a}. 
These voids exhibit some intriguing properties.
For example, while apparently just simple vacant spaces, they actually contain
a complex, multi-level hierarchical 
dark matter substructure~\citep{vandeWey1993,Gottlober2003,Aragon2012}.
Indeed, the interiors of voids appear as miniature cosmic 
webs, albeit at a different mean density~\citep{Goldberg2004}.
However, these void substructures obey simple scaling relations 
that enable direct translations 
of void properties between different tracer types (e.g., 
galaxies and dark matter)~\citep{Benson2003,Ricci2014,Sutter2013a}.
Their internal growth is relatively simple: voids 
do not experience major merger events over their long 
lifetimes~\citep{Sutter2014a} and 
their interiors are largely unaffected by
larger-scale environments~\citep{Dubinski1993,Fillmore1984,vandeWey2009}.

As underdense regions, voids are the first objects in the large-scale 
structure to be dominated by dark energy. This fact coupled with their 
simple dynamics makes them a unique and potentially potent 
probe of cosmological parameters, either through their 
intrinsic properties~\citep[e.g.,][]{Biswas2010,Bos2012}, 
exploitation of their statistical isotropy via the 
\ap~test~\citep{LavauxGuilhem2011,Sutter2012b,Sutter2014b}, 
or cross-correlation with the cosmic microwave 
background \citep{Thompson1987, Granett2008, Ilic2013, Planck2013b, Cai2014}.
Additionally, fifth forces and modified gravity are unscreened 
in void environments, making them singular probes 
of exotic 
physics~\citep[e.g.,][]{Li2012,Clampitt2013,Spolyar2013,Carlesi2013a}.

Voids offer a unique laboratory for studying the relationship 
between galaxies and dark matter unaffected by complicated 
baryonic physics. As noted above, there appears to be 
a self-similar relationship between voids in dark 
matter distributions and voids in galaxies~\citep{Sutter2013a} 
via a universal density 
profile~\citep[][hereafter HSW]{Hamaus2014}.
Void-galaxy cross-correlation analyses also reveal a striking 
feature: the large-scale clustering power of compensated voids 
is identically zero, 
which may give rise to a static cosmological ruler~\citep{Hamaus2013}.
Observationally, measurements of the anti-lensing shear signal 
of background galaxies have revealed the internal dark matter 
substructure in voids~\citep{Melchior2014,Clampitt2014}, and 
Ly-alpha absorption measurements have illuminated dark matter 
properties in void outskirts~\citep{Tejos2012}.
Finally, studying the formation of void galaxies reveals the 
secular evolution of dark matter halos~\citep{Rieder2013} 
and their mass function~\citep{Neyrinck2014}.

Voids present a useful region for investigating astrophysical 
phenomena, as well. For example, the detection of magnetic
fields within voids constrains the physics of the primordial 
Universe~\citep{Taylor2011,Beck2013}. Contrasting galaxies in 
low- and high-density environments probes the relationship 
between dark matter halo mass and galaxy 
evolution~\citep{VandeWeygaertR.2011, Kreckel2011, Ceccarelli2012, Hoyle2012}.

Given the burgeoning interest in voids, there remains surprisingly
little publicly-accessible void information.
There are a few public catalogs of voids identified 
in galaxy redshift surveys, primarily the 
SDSS~\citep[e.g.,][]{Pan2011, Sutter2012a, Nadathur2014, 
Sutter2013c, Sutter2014b}, and there are fewer still catalogs 
of voids found in simulations and mock galaxy populations~\citep{Sutter2013a}.
And while there are many published methods for finding voids based 
on a variety of techniques, such as spherical underdensities
\citep{Hoyle2004,Padilla2005}, watersheds \citep{Platen2007,Neyrinck2008}, 
and phase-space dynamics~\citep{Lavaux2010,vandeWey2010,Sousbie2011,Cautun2013,Neyrinck2013}, 
most codes remain private. 
In order to accommodate the expanding application of voids and to engender 
the development of communities and collaborations, it is essential 
to provide easy-to-use, flexible, and strongly supported 
void-finding codes.

In this paper we present \vide\footnote{The French word for ``empty''.},
for Void IDentification and Examination, 
a toolkit based on the publicly-available 
watershed code \textsc{zobov}~\citep{Neyrinck2008} for finding voids 
but considerably enhanced and extended to handle a variety of simulations 
and observations. \vide~also provides an extensive interface for 
analyzing void catalogs. In Section~\ref{sec:inputs} we outline the 
input data options for void finding, followed by Section~\ref{sec:finding} 
where we describe our void finding technique and 
our extensions and modifications to \textsc{zobov}.
Section~\ref{sec:analysis} details our Python-based analysis toolkit 
functionality, and Section~\ref{sec:guide} is a quick-start user's 
guide. We summarize and provide an outlook 
for future uses and upgrades to \vide~in Section~\ref{sec:conclusions}.

\section{Input Data Options}
\label{sec:inputs}

\subsection{Simulations}

To identify voids in $N$-body dark matter populations,
\vide~is able to read {\tt Gadget}~\citep{Gadget},
{\tt FLASH}~\citep{Dubey2008}, and {\tt RAMSES}~\citep{ramses} 
simulation outputs, 
files in the Self-Describing Format~\citep{warren2013}, and generic ASCII 
files listing positions and velocities.
Void finding can be done on the dark matter particles themselves, 
or in randomly subsampled subsets with user-defined mean densities.
Subsampling can be done either in a post-processing step or 
\emph{in situ} during void finding.

\vide~can also find voids in halo populations. The user must provide 
an ASCII file and specify the columns containing the halo mass, 
position, and other properties. The user can use all identified 
halos or specify a minimum mass threshold for inclusion in the 
void finding process.

The user may construct a mock galaxy population from  a halo 
catalog using a Halo Occupation
Distribution (HOD) formalism~\citep{Berlind2002}. 
HOD modeling assigns central and satellite galaxies to a dark matter
halo of mass $M$ according to a parametrized distribution.
\vide~implements the 
five-parameter model of~\citet{Zheng2007}, where
the
mean number of central galaxies is given by
\begin{equation}
\left\langle N_{\rm cen}(M)\right\rangle = \frac{1}{2} \left[
1 + \erf \left(\frac{\log M - \log M_{\rm min}}{\sigma_{\log M}}\right)
\right]
\end{equation}
and the mean number of satellites is given by
\begin{equation}
\left< N_{\rm sat}(M)\right> = \left\langle N_{\rm cen}(M) \right\rangle
\left( \frac{M-M_0}{M_1'}\right)^\alpha,
\end{equation}
where $M_{\rm min}$, $\sigma_{\log M}$, $M_0$, $M_1'$, and $\alpha$
are free parameters that must be fitted to a given survey.
The probability distribution of central galaxies is a nearest-integer
distribution (i.e., all halos above a given mass threshold host a central
galaxy), and satellites follow Poisson statistics,
These satellites 
are placed around the central galaxy with random positions 
 assuming a probability given by the NFW~\citep{NFW} profile for a halo of the given mass.
The user can also specify an overall mean density in case the HOD 
model was generated from a simulation with different cosmological 
parameters than the one used for void finding, which causes 
a mismatch in the target galaxy density. While not a full fix (which 
would require a new HOD fit), this at least alleviates some of 
the mismatch.

We have included --- but not fully integrated into the pipeline --- 
a separate code for fitting HOD parameters to a given simulation.
The implemented model has three parameters, and given two of those 
the third is fixed by demanding that the abundance of galaxies of a given 
population reproduces the observed value. We then explore the 2-dimensional 
space of the other two parameters trying to minimize the $\chi^2$ that 
results from comparing the two-point correlation function of the simulated 
galaxies with the observed data.

In all cases (particles, halos, or mock galaxies), void finding can be done 
on the real-space particle positions or after being placed on 
a lightcone (i.e., redshift space) assuming given cosmological 
parameters, where the $z$-axis of the 
simulation is taken to be the line of sight. 
Voids can also be found after particle positions have been perturbed 
using their peculiar velocities.
The user may define independent slices and sub-boxes of the full simulation 
volume, and \vide~will handle all periodic effects and 
boundary cleaning (see Section~\ref{sec:finding}) automatically.

\subsection{Observations}

\begin{figure}  
  \centering 
  {\includegraphics[type=png,ext=.png,read=.png,width=\columnwidth]{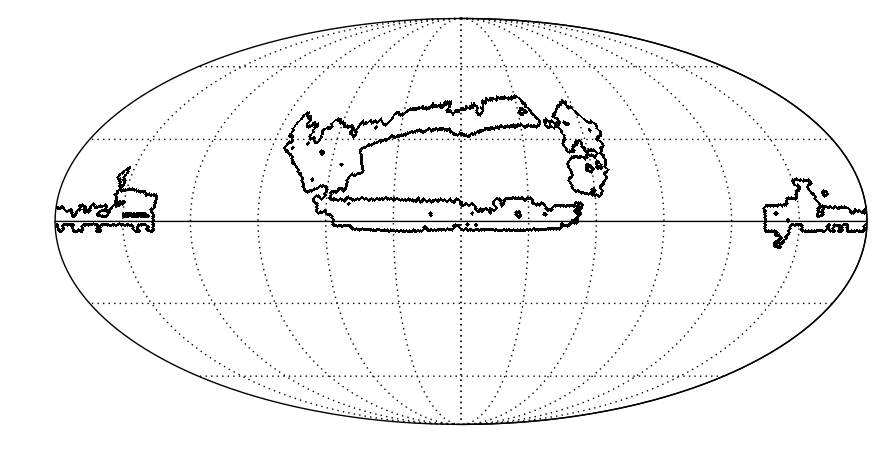}}
  \caption{ Example \textsc{healpix} map in a Mollweide projection
                  of identified boundary zones (black) around and
                  within the SDSS DR9 survey area. Maps like this
                  are used to prevent voids from being identified outside 
                  a survey volume.
            \emph{Reproduced from~\citet{Sutter2013c}}.
          }
\label{fig:mask}
\end{figure}

In addition to providing an ASCII file listing galaxy right ascension, 
declination, and redshift, the user must provide a 
pixelization of the survey mask
using \textsc{healpix} \citep{Gorski2005}\footnote{http://healpix.jpl.nasa.gov}.
The \textsc{healpix} description of the sphere provides
equal-area pixels, and the \textsc{healpix} implementation
itself provides built-in tools to easily
determine which pixels lie
on the boundary between the survey area and any masked region.
This is essential for \vide~to constrain voids to the survey volume 
(see Section~\ref{sec:finding}).
Figure~\ref{fig:mask} shows a pixelization of the SDSS DR9 mask
and the location of the boundary pixels. To accurately capture
the shape of the mask we require a resolution of at least 
$n_{\rm side}=512$.

\vide~also provides a utility for constructing a rudimentary mask from the 
galaxy positions themselves.

The default \vide~behavior is to assume that the galaxy population 
is volume-limited (i.e., uniform galaxy density as a function of redshift). 
However, the user may provide a selection function and identify 
voids in a full magnitude-limited survey. As mentioned in 
~\citet{Neyrinck2008}, in this case the
Voronoi densities will be weighted prior to void finding such that
$\rho' = \rho/w(z)$, where $w(z)$ is the relative number density as a function 
of redshift (note that this functionality can also be used for arbitrary 
re-weighting, if desired). This re-weighting is only used 
in the watershed phase for purposes of constructing voids, and does 
not enter into later volume or density calculations.
If re-weighting is chosen then the 
selection function will be used to calculate a global mean density.

\section{Void Finding}
\label{sec:finding}

The core of our void finding algorithm is
\textsc{zobov}~\citep{Neyrinck2008}, 
which creates a Voronoi tessellation of the tracer particle
population and uses the watershed transform to group Voronoi
cells into zones and subsequently voids~\citep{Platen2007}.
The Voronoi tessellation provides a density field estimator 
based on the underlying particle positions.
By implicitly
performing a Delaunay triangulation (the dual of the Voronoi
tessellation), \textsc{zobov} assumes constant density across the volume
of each Voronoi cell, which effectively 
sets a local smoothing scale for the continuous
field necessary to perform the
watershed transform. There is no additional smoothing.
For magnitude-limited surveys, where the mean galaxy density 
varies as a function of redshift, we weight the Voronoi cell volumes 
by the local value of the radial selection function.

To construct voids
the algorithm first groups nearby Voronoi
cells into {\emph zones}, which are local catchment basins.
Next, the watershed transform merges adjacent zones into voids
by finding minimum-density barriers between them and joining zones
together to form larger agglomerations.
We impose a density-based threshold within \textsc{zobov} where
adjacent zones are only added to a void
if the density of the wall between them is
less than $0.2$ times the mean particle density $\bar{n}$.
For volume-limited surveys we calculate the mean particle density by 
dividing the survey volume (from the \textsc{healpix} 
mask and redshift extents) 
by the total number of galaxies. 
For magnitude-limited surveys the mean particle density is estimated from the 
selection function.
This density criterion 
prevents voids from expanding deeply into overdense structures and
limits the depth of the void hierarchy~\citep{Neyrinck2008}.

However, this process does not place a restriction on the density of the
initial zone, and thus in principle a void can have any minimum density.
By default \vide~reports every identified basin as a void, but in the 
section below we describe some provided facilities for 
filtering the catalog based on various criteria, depending on the 
specific user application.

\begin{figure}
  \centering
  {\includegraphics[type=png,ext=.png,read=.png,width=\columnwidth]{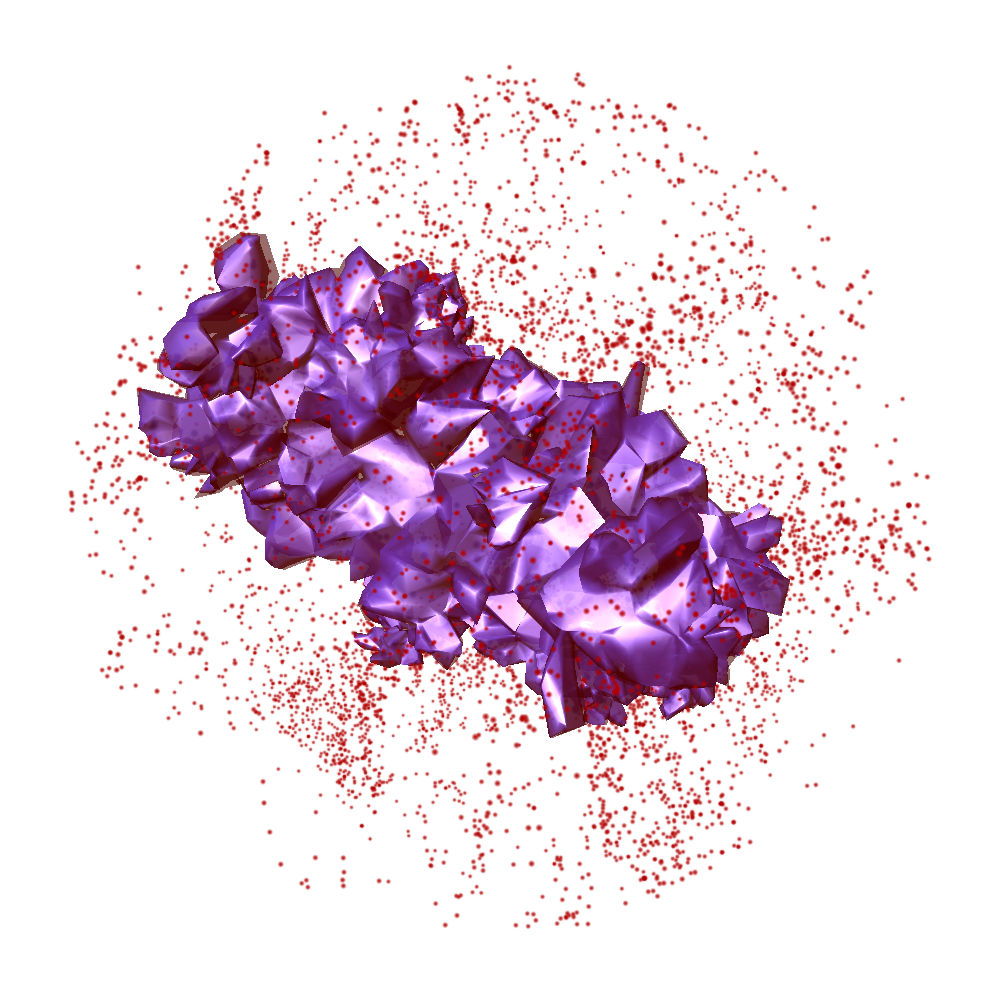}}
  \caption{An example of a watershed void in \vide. 
           The Voronoi cells that define the void are in purple
           with galaxies in red. We show a void with effective
           radius 20~\hmpc~within
           a 50~\hmpc~spherical region.
           Galaxy point sizes are proportional
           to their distance from the point of view. Galaxies
           interior to the void are shaded dark red.
            \emph{Reproduced from~\citet{Sutter2012a}}.
           }
\label{fig:void}
\end{figure}

In this picture, a void is simply a depression in the density field:
voids are
aspherical aggregations of Voronoi cells that share a common
low-density basin and are bounded by a common set of higher-density walls,
as demonstrated by Figure~\ref{fig:void}, which shows a typical 
20 \hmpc~void identified in the SDSS DR7 galaxy survey~\citep{Sutter2012a}.
This also means that voids may have any \emph{mean} density, since 
the watershed includes in the void definition all wall particles 
all the way up to the very highest-density separating ridgeline.

\begin{figure} 
  \centering  
  {\includegraphics[type=pdf,ext=.png,read=.png,width=\columnwidth]{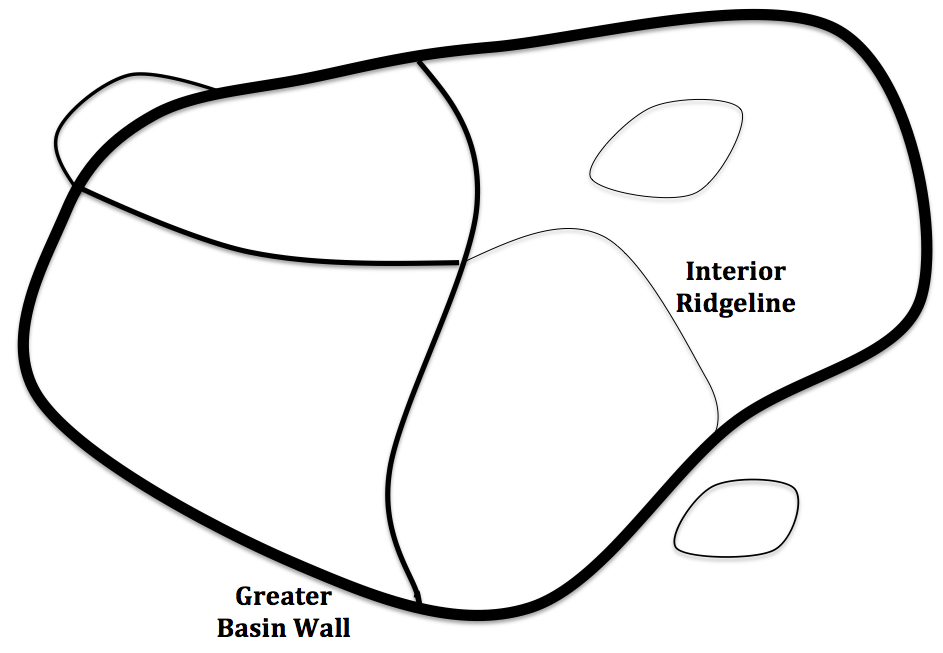}}
  {\includegraphics[type=pdf,ext=.pdf,read=.pdf,width=\columnwidth]{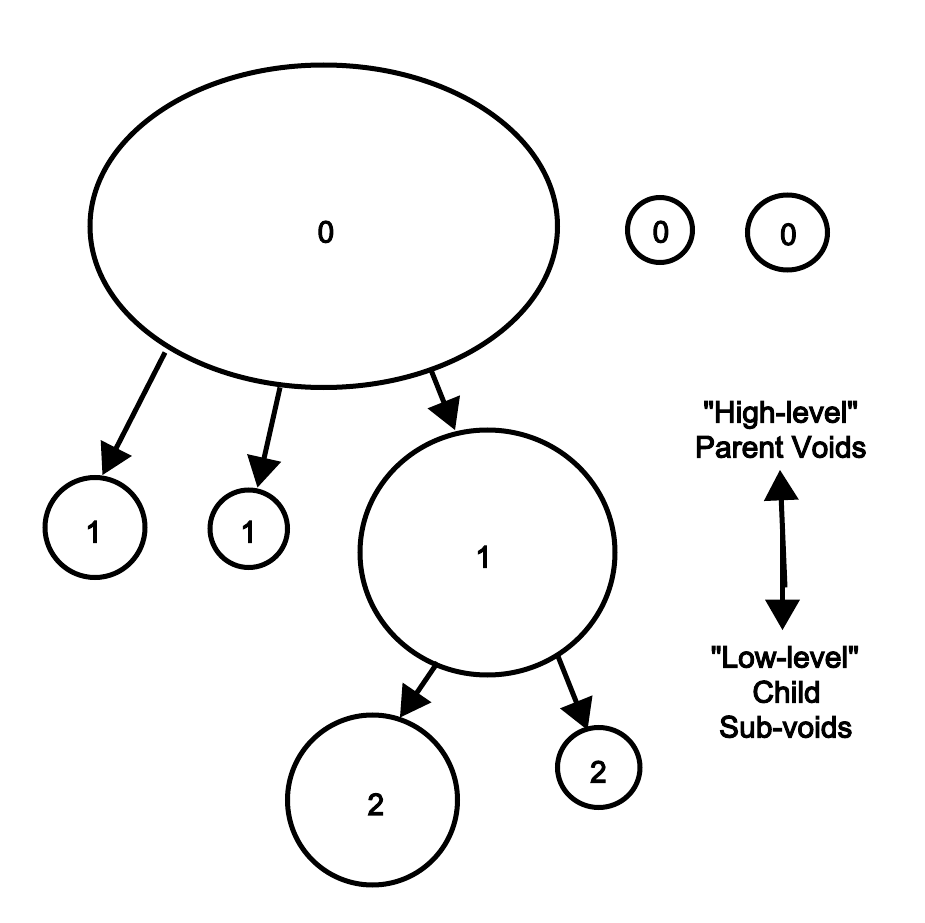}}
  \caption{
           A cartoon of the assembly of the void hierarchy. The top panel
           shows ridgelines with line thickness proportional to density.
           The bottom panel shows the tree derived from such a collection
           of voids, with the tree level of each void indicated.
            \emph{Reproduced from~\citet{Sutter2014a}}.
          }
\label{fig:hierarchy}
\end{figure}

We may construct a nested hierarchy of
voids~\citep{LavauxGuilhem2011, Bos2012} using the topologically-identified 
watershed basins and ridgelines.
We begin by identifying the initial
zones as the deepest voids, and as we progressively merge voids
across ridgelines we establish super-voids. There is no unique
definition of a void hierarchy, and we take the semantics
of~\citet{LavauxGuilhem2011}: a parent void contains all the zones
of a sub-void plus at least one more. All voids have only one
parent but potentially many children, and the children
of a parent occupy distinct subvolumes separated by low-lying
ridgelines. There are also childless ``field'' voids. 
Figure~\ref{fig:hierarchy} shows a cartoon of this
void hierarchy construction.
Without the application of the $0.2 \bar{n}$ density cut 
discussed above, \textsc{zobov} would identify a single super-void 
spanning the entire volume, and thus there would be a single 
hierarchical tree. However, with the cut applied there are multiple 
top-level voids.

For historical reasons~\citep{Sutter2012b,Sutter2012a}, the default 
catalog removes voids with high central densities 
($\rho (R < 0.25 R_{\rm eff}) > 0.2 \bar \rho$) and 
any children voids, but the 
user can trivially access all voids.

We have made several modifications and improvements to the original
\textsc{zobov} algorithm, which itself is an inversion of the 
halo-finding algorithm \textsc{voboz}~\citep{voboz}. 
First, we have made many speed and performance 
enhancements, as well as re-written large portions in a 
templated C++ framework for more modularity and flexibility.
This strengthens
\textsc{zobov} with respect to numerical precision.
Floating-point precision can occasionally lead to disjoint 
Voronoi graphs, especially in high-density regions: one particle
may be linked to another while its partner does not link back to it. 
We therefor enforce bijectivity in the Voronoi graph 
(so that the tessellation is self-consistent) by ensuring that all 
links are bidirectional. 

We apply \textsc{zobov} to very large simulations. To run the analysis on such
simulations while keeping memory consumption within reasonable bounds, it is necessary to split the volume on which the Delaunay
tesselation is done into several subvolumes. Additionally many large
simulations store the particle positions in single precision mode. The original
\textsc{zobov} re-shifted each subvolume such that their geometrical center
was always at the coordinate $(0,0,0)$. 
However that involves computing differences with
single precision floats, increasing the numerical noise. We found
practically that shifting the subvolume 
to place a corner at $(0,0,0)$ incurs slightly fewer problems. 
However, this is insufficient to completely mitigate the
issue that when subvolumes are merged the volumes of the tetrahedra are
not exactly the same on both sides of the subvolume boundary. So we
again enforce the connectivity of the mesh, but lose the Delaunay
properties at each subvolume boundary. 
While this is not exact, it still constructs
a fully connected mesh and allows the computation of the watershed transform.

Finally, we have improved the performance of the watershed transform in
\textsc{zobov} using a priority queue algorithm to sort out the zones that need
to be processed according to their core density and spilling 
density (the saddle point with the minimum density).
The priority queue algorithm has a $\mathcal{O}(1)$ 
time complexity for getting the
next element and replaces the $\mathcal{O}(N)$ full 
search algorithm from the original \textsc{zobov}. 
The cost of insertion is at most the number
of zones that were not processed but in practice this is negligible.
With these improvements we have identified voids in simulations 
with up to $1024^3$ particles in $\sim 10$ hours using 
16 cores.

By default \vide~also use the latest version of the {\tt qhull}
library\footnote{\mbox{http://www.qhull.org}}~\citep{qhull}, where we take 
advantage of provided functions 
for constructing Voronoi graphs that are more stable in high-density 
regions.

Since \textsc{zobov} was originally developed in the context of 
simulations with periodic boundary conditions, care must be taken 
in observations and simulation sub-volumes.
To prevent the growth of voids outside survey boundaries,
we place a large number of mock particles
along any identified edge (Figure~\ref{fig:mask}) and along 
the redshift caps. 
The user inputs the density of mock particles, and \vide~places 
them randomly within the projected volume of the 
\healpix~pixels. To prevent spillover and misidentification 
of edge zones, studies indicate that the density of these 
mock particles should be at least the density of the 
sample studied, and preferably as high as computationally 
tolerable~\citep{Sutter2012a, Nadathur2014, Sutter2013d}.
We assign essentially infinite density to these mock particles, 
preventing the watershed procedure from merging zones external 
to the survey.
Since their local volumes are arbitrary, 
we prevent these mock particles from participating in any 
void by disconnecting their adjacency links in the Voronoi graph.

This process leaves a population of voids near --- but not directly 
adjacent to --- the survey edge, which can induce a subtle 
bias in the alignment distribution, affecting 
cosmological probes such as the \ap~test~\citep{Sutter2012b}. 
Also, while these \emph{edge} voids are indeed underdensities, 
making them useful for certain applications, their 
true extent and shape is unknown. \vide~thus provides 
two void catalogs: \emph{all}, which includes every identified 
void, and \emph{central},
where voids are guaranteed to sit well
away from any survey boundaries or internal holes: the maximum
distance from the void center to any member particle is less than the
distance to the nearest boundary. 
Voids near any redshift caps (defined using the same criterion)
are removed from all catalogs.

For subvolumes taken from a larger simulation box (see the 
discussion above for input data options), the edges 
of the subvolume are assumed to be periodic for 
purposes of the watershed, regardless of whether that edge is actually 
periodic or not. However, any void near a non-periodic 
edge is removed from all catalogs using the distance criterion 
described in the preceding paragraph, since these voids will 
have ill-defined extents. For all simulation-based 
catalogs there is no difference between the \emph{all} and 
\emph{central} catalogs.

We use the mean particle spacing to set a lower size limit for voids
because of the effects of shot noise. \vide~does not include any void with 
effective radius smaller than $\bar{n}^{-1/3}$, where $\bar{n}$ is the 
mean number density of the sample.
We define the effective radius as
\begin{equation}
  R_{\rm eff} \equiv \left( \frac{3}{4 \pi} V \right)^{1/3},
\end {equation}
where $V$ is the total volume of all the Voronoi cells that make up the void.

\begin{figure}
  \centering
  {\includegraphics[type=pdf,ext=.pdf,read=.pdf,width=\columnwidth]{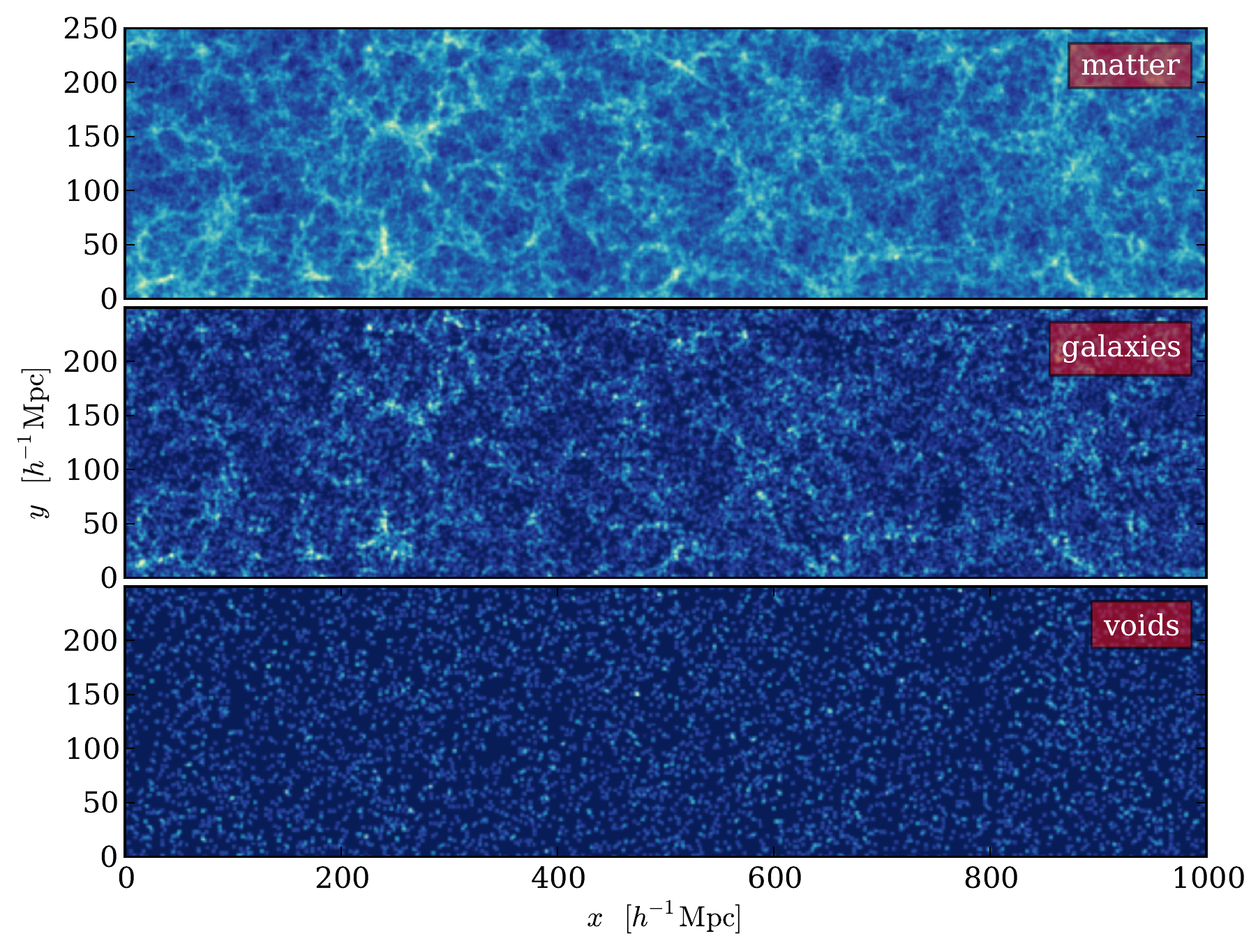}}
  \caption{
           Projected density fields of dark matter (top), mock galaxies (middle) and voids (bottom) in a $250$\hmpc~slice of a simulation box. 
            \emph{Reproduced from~\citet{Hamaus2013}}.
  }
  \label{fig:density}
\end{figure}

Figure~\ref{fig:density} shows an example void population with
\vide, taken from 
the analysis of~\citet{Hamaus2013}. In this figure we show a slice 
from an $N$-body simulation, a set of mock galaxies painted onto the 
simulation using the HOD formalism discussed above, and the distribution 
of voids identified in the mock galaxies.

\vide~provides some basic derived void information.
The most important quantity is the \emph{macrocenter}, or volume-weighted
center of all the Voronoi cells in the void:
\begin{equation}
  {\bf X}_v = \frac{1}{\sum_i V_i} \sum_i {\bf x}_i V_i,
\label{eq:macrocenter}
\end{equation}
where ${\bf x}_i$ and $V_i$ are the positions and Voronoi volumes of
each tracer particle $i$, respectively.
\vide~also computes void shapes by taking void member particles 
and constructing the inertia tensor:
\begin{eqnarray}
  M_{xx} & = &\sum_{i=1}^{N_p} (y_i^2 + z_i^2) \\ 
  M_{xy} & = & - \sum_{i=1}^{N_p} x_i y_i, \nonumber
\end{eqnarray}
where $N_p$ is the number of particles in the void, and
$x_i$, $y_i$, and $z_i$ are coordinates of the particle $i$
relative to the void macrocenter.
The other components of the tensor are obtained by 
cyclic permutations.
This definition of the inertia tensor is designed to give greater weight 
to the particles near the void edge, which is useful for applications 
such as the \ap~test. Other definitions, such as volume-weighted 
tensors, can be implemented trivially with the toolkit functionality
 discussed below.
We use the inertia tensor to compute eigenvalues and 
eigenvectors and form the ellipticity:
\begin{equation}
  \epsilon = 1- \left( \frac{J_1}{J_3}\right)^{1/4},
\label{eq:ellip}
\end{equation}
where $J_1$ and $J_3$ are the smallest and largest eigenvalues 
of the inertia tensor, respectively.

\section{Post-Processing \& Analysis}
\label{sec:analysis}

\vide~provides a Python-based application programming interface (API) 
for loading and manipulating the void catalog and performing 
analysis and plotting.
The utilities described below present simulation and observation
void catalogs on an equal footing: density and volume normalizations, the 
presence of boundary particles, and other differences are handled 
internally by \vide~such that the user does not need to implement 
special cases, and simulations and observations may be directly compared 
to each other.

All the following analysis routines are compatible with
releases of the Public Cosmic Void Catalog after version \emph{2013.10.25}.
Complete documentation is available at \linebreak\mbox{http://bitbucket.org/cosmicvoids/vide\_public/wiki/}.

\subsection{Catalog Access}

After loading a void catalog, the user has immediate access to all 
void properties (ID, macrocenter, radius, density contrast, RA, Dec, hierarchy 
level, ellipticity, etc.)
as well as the positions ($x$, $y$, $z$, RA, Dec, redshift),
velocities (if known), and local Voronoi volumes of all void member particles. 
By default these quantities are presented to the user as members of 
an object, but we provide a utility for selectively converting these
quantities to NumPy arrays for more efficient processing.
The user additionally has access to all particle or galaxy sample 
information, such as redshift extents, the mask, simulation extents 
and cosmological parameters, 
and other information.
Upon request the user can also load all particles in the simulation 
or observation. 

Since by default \vide~returns every void above the minimum size 
threshold set by the mean interparticle spacing of the sample, 
we provide some simple facilities for performing the most common 
catalog filtering. For example, the user may select voids 
based on size, position in the hierarchy, central density, 
minimum density, and density contrast. 

\subsection{Plotting}

\begin{figure} 
  \centering 
  {\includegraphics[type=png,ext=.png,read=.png,width=\columnwidth]{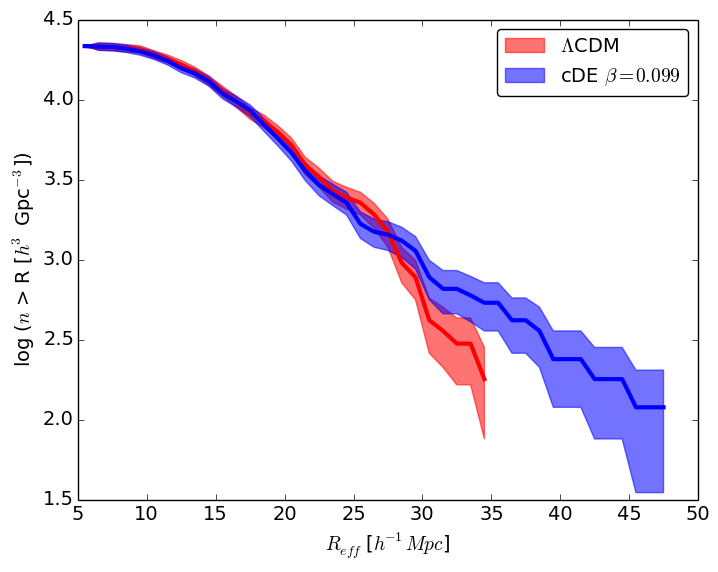}}
  \caption{Example cumulative void number functions from simulations.
           Shown are abundances for \lcdm~(red) and a coupled dark matter-dark 
           energy model (blue).
           The solid lines are the measured number functions and the
           shaded regions are the 1$\sigma$ Poisson uncertainties.
           \emph{Reproduced from~\citet{Sutter2014c}}.
           }
\label{fig:numberfunc}
\end{figure}

\vide~includes several built-in plotting routines.
First, users may plot cumulative number functions of multiple 
catalogs on a logarithmic scale. 
Volume normalizations are handled automatically, 
and 1$\sigma$ Poisson uncertainties are shown as shaded regions, 
as demonstrated in Figure~\ref{fig:numberfunc}. 
Secondly, users may plot a slice from a single void and its 
surrounding environment.
In these plots we bin the background particles onto a 
two-dimensional grid and plot the density on a logarithmic scale. 
We draw the void member galaxies as small semi-transparent disks with
radii equal to the effective radii of their corresponding Voronoi
cells. 
Figure~\ref{fig:matching} highlights these kinds of plots.
The user may also plot an ellipticity distribution for 
any sample of voids.
All plots are saved in png, pdf, and eps formats.

\subsection{Catalog Comparison}

The user can directly compare two void catalogs by using a built-in 
function for computing the overlap. This function attempts to 
find a one-to-one match between voids in one catalog and another 
and reports the relative properties (size, ellipticity, etc.) 
between them. It also records which voids cannot find a reliable match 
and how many potential matches a void may have. 

The potential matches are found by considering all voids in the
matched void catalog whose centers lie within the watershed volume of the
original void. 
Then for each potential matched void the user can choose 
to use either the unique particle IDs or 
the amount of volume overlap to find matches.
We take the potential matched void with the greatest amount of
overlap (volume or number of particles) as the best match.

In the case of volume overlap, to simplify the measurement 
we place each particle at the center
of a sphere whose volume is the same as its Voronoi cell. 
We approximate the Voronoi volumes as spheres to provide a 
stricter definition of overlap, since the Voronoi volume 
can be highly elongated and lead to unwanted matching.
We
measure the distance between particles and assume they overlap if
their distances meet the criterion
\begin{equation}
  d \le \alpha ( R_1 + R_2 ),
\end{equation}
where $d$ is the distance and $R_1$ and $R_2$ are the radii of the spheres
assigned to particles 1 and 2, respectively. 
The user may select the value $\alpha$, but
we found the factor
of $\alpha=0.25$ to strike the best balance between conservatively estimating
overlap while still accounting for our spherical estimate
of the Voronoi volume of each particle. 
If the particles meet the distance criterion, the volume is added 
to the total amount of overlap; the void with the most amount of overlap 
is considered the best match.

\begin{figure} 
  \centering 
  {\includegraphics[type=pdf,ext=.pdf,read=.pdf,width=\columnwidth]{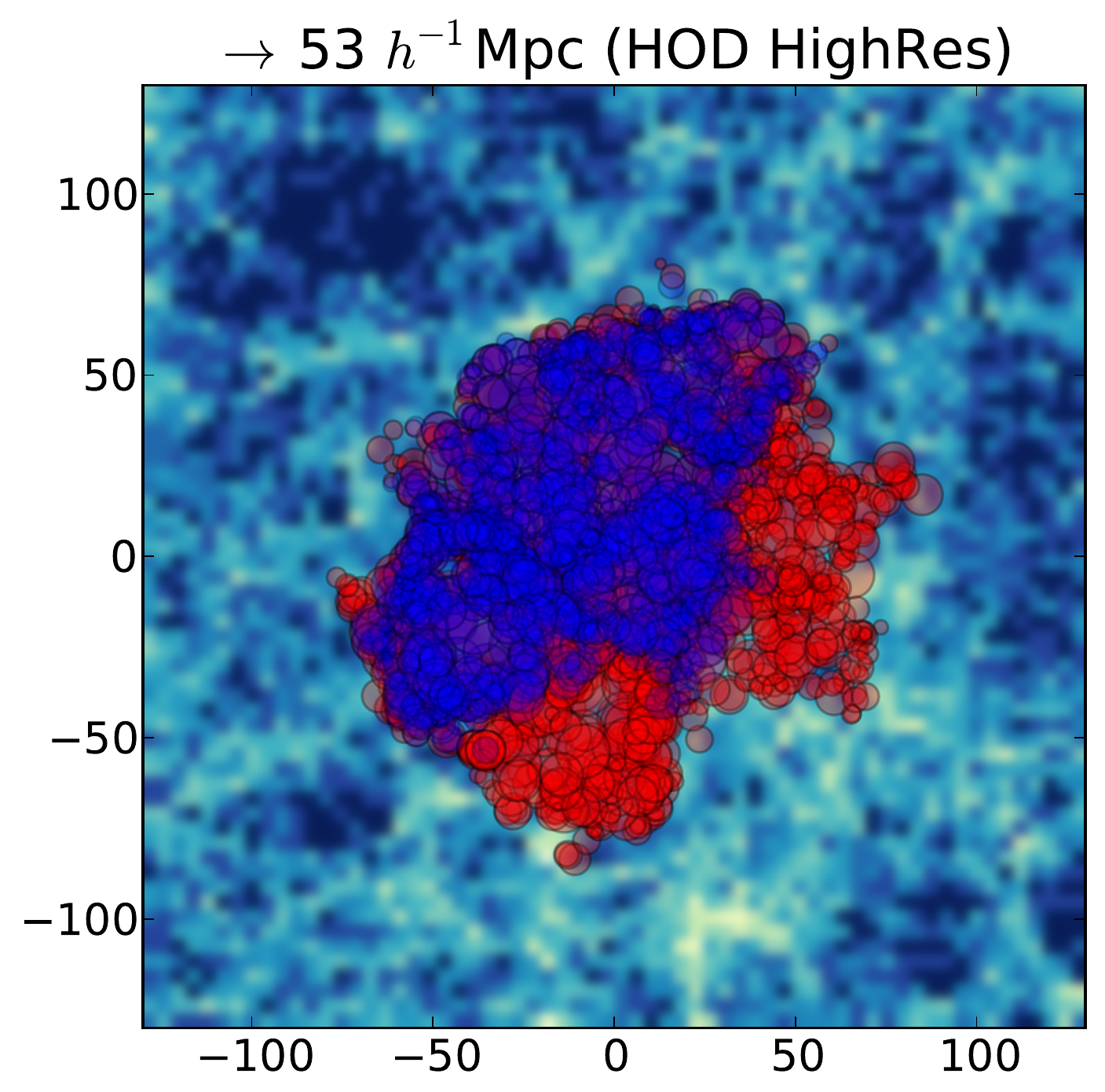}}
  \caption{ Example of matching and void plotting. Here a void identified 
            in real space (blue circles) has been matched to a void identified 
            in redshift space (red circles) in an HOD mock galaxy 
            population.  
            The background particles are binned and the corresponding 
            density is shown on a logarithmic scale from $0.0$ (white) 
            to $1.5$ (black).
            The void member galaxies are shown as small 
            semi-transparent disks with
            radii equal to the effective radii of their corresponding Voronoi
            cells. 
            In this case the width of the slice is 50 \hmpc.
            \emph{Reproduced from~\citet{Pisani2014}}.
           }
\label{fig:matching}
\end{figure}

A more detailed discussion of the matching process can be 
found in~\citet{Sutter2013b}.
Figure~\ref{fig:matching} shows how a void identified in a real-space 
HOD mock galaxy population has been matched to a void identified 
in a corresponding redshift-space population.

\subsection{Clustering Statistics}
\vide~allows the user to compute simple two-point clustering statistics, i.e. power spectra and correlation functions. To perform this, \vide~reads in particle positions and uses a cloud-in-cell mesh interpolation scheme~\citep{Hockney1988} to construct a three-dimensional density field of fluctuations
\begin{equation}
\delta(\mathbf{r}) = \frac{n(\mathbf{r})}{\bar{n}} - 1\;,
\end{equation}
where $n(\mathbf{r})$ is the spatially varying number density of tracers with a mean of $\bar{n}$. 

Then, Fourier modes of the density fluctuation with mesh wave vector $\mathbf{k}$ are computed using a standard DFT algorithm (computed using an FFT),
\begin{equation}
\delta(\mathbf{k})=\frac{1}{N_c}\sum_\mathbf{r} \delta(\mathbf{r})\exp(-i\mathbf{k}\cdot\mathbf{r})\;,
\end{equation}
and the angle-averaged power spectrum is estimated as
\begin{equation}
P(k) = \frac{V}{N_k}\sum_{\Delta k}\frac{\left|\delta(\mathbf{k})\right|^2}{W^2_\mathrm{cic}(\mathbf{k})}\;.
\end{equation}
Here $V$ denotes the simulation volume, $N_c$ the number of grid cells in the mesh, $N_k$ the number of Fourier modes in a $k$-shell of thickness $\Delta k$, and
\begin{equation}
W_\mathrm{cic}(\mathbf{k})=\prod_{i=x,y,z}\frac{\sin^2(k_i)}{k_i^2} \;
\end{equation}
is the cloud-in-cell window function to correct for artificial suppression of power originating from the mesh assignment scheme. An inverse Fourier transform of the power spectrum yields the correlation function,
\begin{equation}
\xi(r)=\frac{1}{V}\sum_\mathbf{k} P(\mathbf{k})\exp(i\mathbf{k}\cdot\mathbf{r})\;.
\end{equation}

\begin{figure*}
\centering
  {\includegraphics[type=pdf,ext=.pdf,read=.pdf,width=0.45\textwidth]{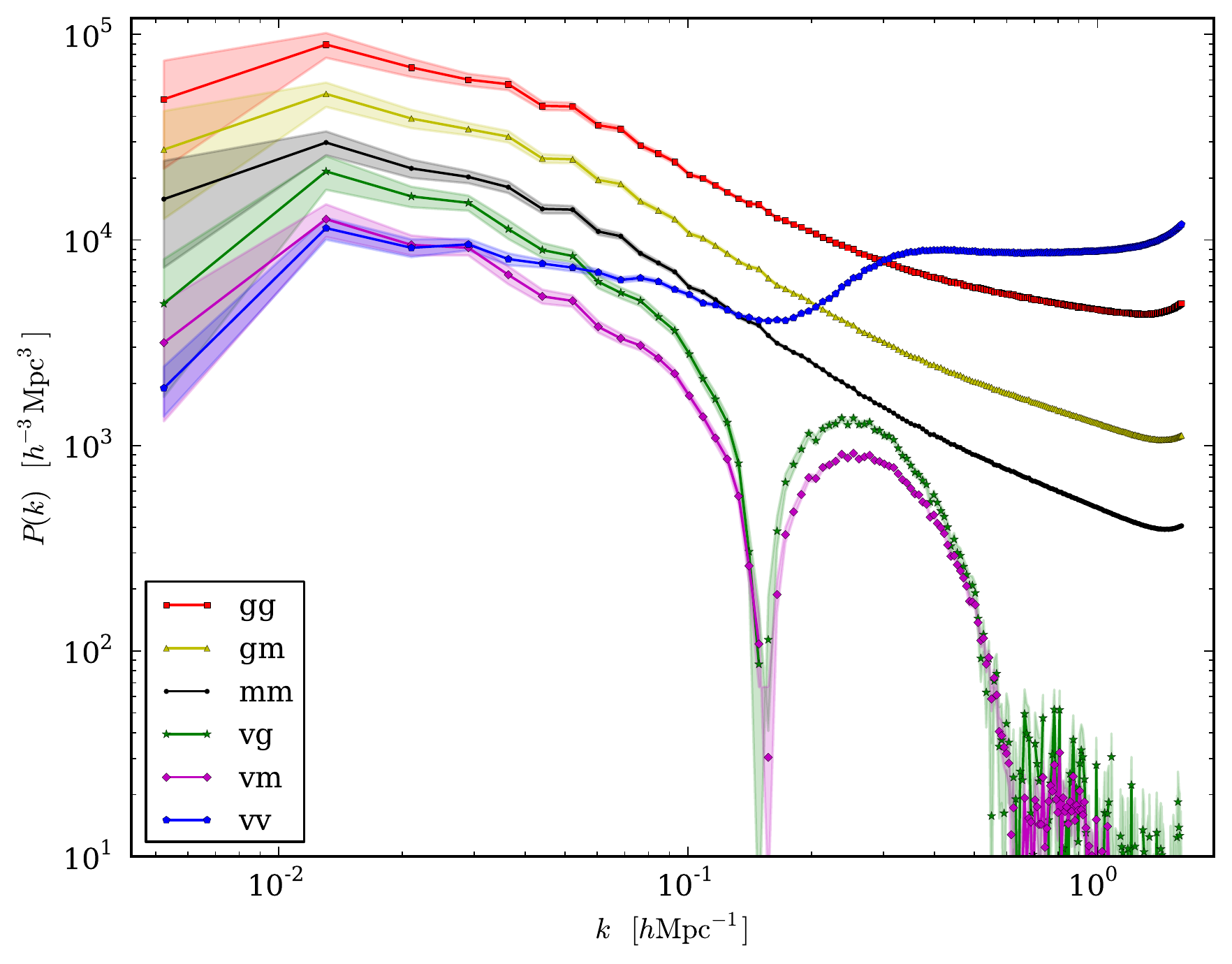}}
  {\includegraphics[type=pdf,ext=.pdf,read=.pdf,width=0.45\textwidth]{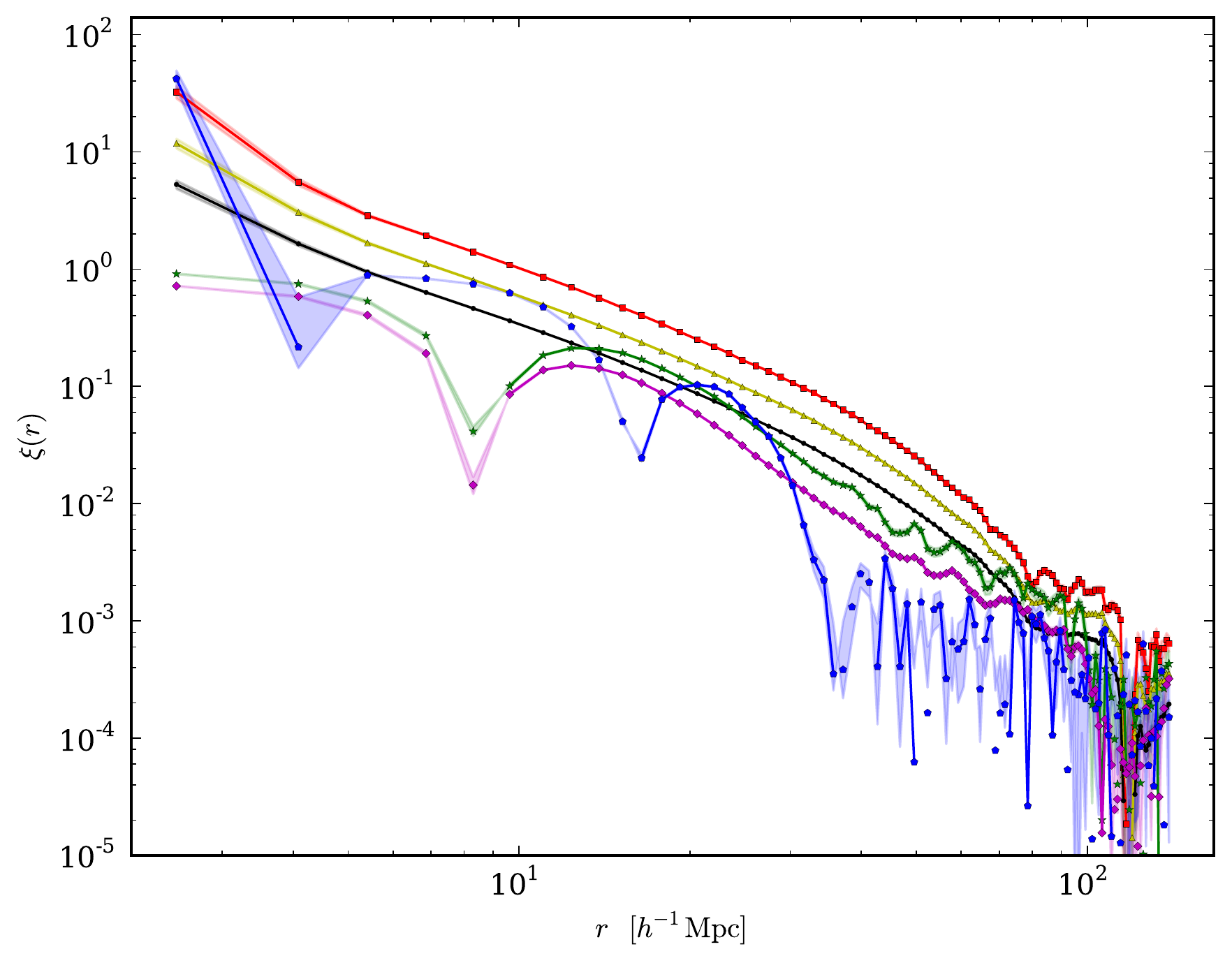}}
  \caption{Example power spectra (left) and correlation 
           functions (right) for various combinations of dark matter particles, 
           mock galaxies, and voids (see inset, solid lines omitted for 
           negative values). 
           Shaded bands show $1\sigma$-uncertainties estimated from scatter 
           in the bin average. 
           \emph{Reproduced from~\cite{Hamaus2013}.}
  }
  \label{fig:xcor1D}
\end{figure*}

As this procedure can be applied to any particle type, \vide~returns power spectra and correlation functions for both void centers and the tracer particles used to identify voids (e.g., dark matter particles or mock galaxies). In addition, it provides the cross-power spectrum and cross-correlation function between the void centers and the tracer particles. This routine creates plots for projected density fields (Figure~\ref{fig:density}) and power spectra and correlation functions (Figure~\ref{fig:xcor1D}).

\subsection{Stacking \& Profiles}

\begin{figure} 
  \centering 
  {\includegraphics[type=png,ext=.png,read=.png,width=\columnwidth]{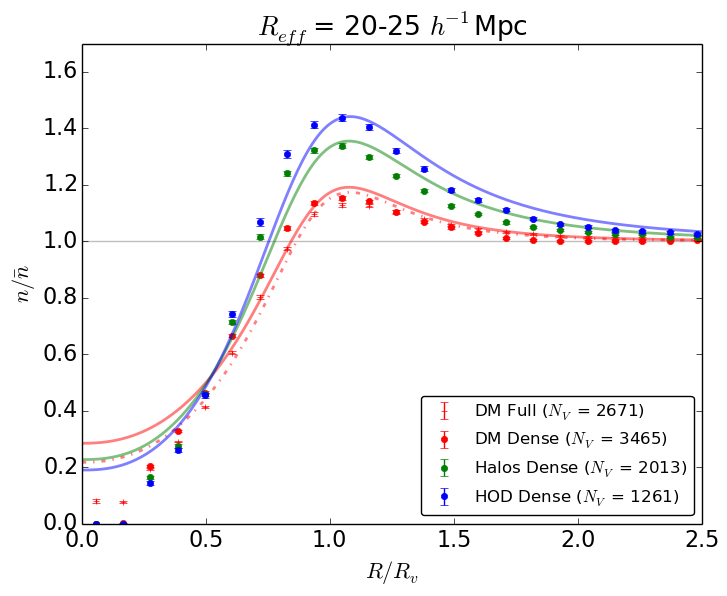}}
  \caption{ 
           Example one-dimensional radial density profiles of stacked voids
           (points with error bars) and best-fit curves (thin lines)
           using the HSW profile.
           Each profile is normalized to the mean number density $\bar{n}$
           of that sample and $R_v$ corresponds to the median
           void size in the stack.
           \emph{Reproduced from~\citet{Sutter2013a}}.
           }
\label{fig:profile}
\end{figure}

The user may construct three-dimensional stacks of voids, 
where void macrocenters are aligned and particle positions are shifted 
to be expressed as relative to the stack center. The stacked void 
may optionally contain only void member particles or all particles 
within a certain radius. Note that this stacking routine does not 
re-align observed voids to share a common line of sight.
With this stacked profile \vide~contains routines for building a 
spherically averaged radial density profile and fitting 
the universal HSW void profile to it. All proper 
normalizations are handled internally.
Figure~\ref{fig:profile} shows an example of \vide-produced 
density profiles and best-fit HSW profile curves. 
These profiles are taken at fixed void size but from many 
different samples, such as high-density dark matter and 
mock galaxy populations.

\vide~also provides routines for reporting theoretical best-fit HSW profiles 
discovered in a variety of populations and void sizes, taken 
from~\citet{Hamaus2014} and~\citet{Sutter2013a}. The user can select 
the sample density, tracer type, and void size closest to their population 
of interest. These profiles are useful for many applications, such as 
ISW predictions and comparing theory to data~\citep[e.g.,][]{Pisani2013}.

\section{Quick-Start Guide}
\label{sec:guide}

\vide~uses Python and C++, and so requires only a few prerequisites prior 
to installation: namely, CMake, Python 2.7, 
NumPy ($\ge$ 1.6.1), and Matplotlib ($\ge$ 1.1.rc). 
\vide~will, by default, automatically 
download and install all other required Python and C++ 
libraries (healpy, cython, GSL, Boost, etc.), unless the 
user indicates via a setup parameter that the
libraries are already available on the system. 

After installation, the user should move to the 
\linebreak \texttt{pipeline/datasets} directory 
and create a \emph{dataset} file. The dataset describes the observation 
or simulation parameters, listing of input files, 
the desired subsampling levels, and various other bookkeeping parameters.
Inputs are prepared with {\tt prepareInputs}, which organizes 
and (if necessary) converts inputs, creates void-finding 
pipeline scripts, and performs subsampling or HOD generation if 
requested by the user.

The user runs the command {\tt generateCatalog} on each pipeline script. 
This command performs the actual void finding in three stages: 
1) further processing of inputs (e.g., application of boundary 
particles) for compatibility with \textsc{zobov}, 2) void finding, 
and 3) filtering of voids near boundaries, construction of the 
void hierarchy, and generation of void property summary outputs.

After void finding \vide~provides a Python 
library \linebreak 
({\tt void\_python\_tools.voidUtil}) for loading, manipulating, and 
analyzing the resulting void catalogs, as presented in the previous 
section. The user simply loads a catalog 
by pointing to an output directory, and has immediate access to all 
void properties and member galaxies. Using this loaded catalog the 
user may exploit all the functionality described above.

Complete documentation is available at \linebreak 
\url{http://bitbucket.org/cosmicvoids/vide\_public/wiki/}.

\section{Conclusions}
\label{sec:conclusions}

We have presented and discussed the capabilities of \vide, a new 
Python/C++ toolkit for identifying cosmic voids in cosmological 
simulations and galaxy redshift surveys. \vide~performs void 
identification using a substantially modified and enhanced version 
of the watershed code \textsc{zobov}. The modifications mostly 
improve the speed and robustness of the original implementation, as
 well as adding extensions to handle observational survey geometries within 
the watershed framework. 
Furthermore, \vide~is able to support a variety of mock and real datasets,
and provides extensive and flexible tools for 
loading and analyzing void catalogs. We have highlighted these 
analysis tools (filtering, plotting, clustering statistics, stacking, 
profile fitting, etc.) using examples from previous and current 
research using \vide.

The analysis toolkit enables a wide variety of both theoretical 
and observational void research. 
For example, the myriad basic and derived void properties available to the 
user, such as sky positions, shapes, and sizes, permit simple explorations 
of void properties and cross-correlation with other datasets.
Extracting void member particles and their associated 
volumes can be used for examining galaxy properties in low-density 
environments. Cross-matching is useful for understanding the 
impacts of peculiar velocities or galaxy bias, as well as providing 
a platform for studying the effects of modified gravity or 
fifth forces on a void-by-void basis. 
Void power spectra, 
shape distributions, number functions, and density profiles, 
easily accessible via \vide, 
are sensitive probes of cosmology. Users may also access 
previously-fit HSW density profiles, enabling theoretical predictions
of the ISW or gravitational lensing signals. 
The ease of filtering the void catalog on various criteria 
allow the user to optimize the selection of voids to maximize 
the scientific return for their particular research aims.

While the current release version of \vide~is fully functional, we do plan a 
number of further enhancements, such as integrating the HOD fitting 
routines directly into the pipeline (rather than the standalone version 
currently packaged), relaxing the \textsc{zobov} constraint of cubic boxes,
and implementing angular selection functions.
In the immediate future
 we will add more plotting functionality
and include two-dimensional clustering statistic construction capabilities. 

\vide~is community-oriented:
the code is currently hosted at \mbox{http://www.bitbucket.org}, which 
supports immediate distribution of updates and bug fixes to the 
user base, as well as providing a host for a wiki 
and bug tracker. Suggestions and fixes will be incorporated into 
the code, with numbered versions serving as milestones. 
Users may utilize the {\tt git} functionality to create branches, 
make modifications, and request merging of their changes into the main 
release.
We have designed \vide~to be as modular as possible: while 
currently based on \textsc{zobov}, any void finder that accepts and outputs 
similar data formats can be included within the toolkit. 

There has been an explosive growth in void interest and research 
in the past decade, as indicated by the number of published void-finding 
algorithms, studies of void properties, and investigations of cosmological 
probes. The release of \vide~is in direct response to the growing demand 
for simple, fast, scalable, and robust tools for finding voids and exploiting 
their properties for scientific gain.
By designing \vide~to be extensible and modular the platform can easily
grow to meet the needs of current and future void communities.

The \vide~code and documentation is currently hosted at \mbox{http://bitbucket.org/cosmicvoids/vide\_public}, with links to numbered versions at \mbox{http://www.cosmicvoids.net}.

\section*{Acknowledgments}

The authors are heavily indebted to Mark Neyrinck, who wisely
and generously made \textsc{zobov} publicly available. 
To recognize that contribution, the authors ask that \textsc{zobov} be cited 
alongside \textsc{vide}.

PMS would like to thank Jeremy Tinker for providing the HOD code used in {\tt VIDE}.

The authors acknowledge
support from NSF Grant NSF AST 09-08693 ARRA. BDW
acknowledges funding from an ANR Chaire d'Excellence (ANR-10-CEXC-004-01),
the UPMC Chaire Internationale in Theoretical Cosmology, and NSF grants AST-0908
902 and AST-0708849.
This work made in the ILP LABEX (under reference ANR-10-LABX-63) was supported by French state funds managed by the ANR within the Investissements d'Avenir programme under reference ANR-11-IDEX-0004-02.

\bibliographystyle{model2-names}
\bibliography{vide}

\end{document}